\documentclass[apj]{emulateapj}
\usepackage{apjfonts}

\defcitealias{loh06}{LS06}
\slugcomment{ApJ, 715, 1486}  

\newcommand\beq{\begin{equation}}
\newcommand\eeq{\end{equation}}

\shorttitle{Statistical Nature of BCGs}
\shortauthors{Lin, Ostriker, \& Miller}

\begin{document}

\title{A New Test of the Statistical Nature of the Brightest Cluster Galaxies}

\author{
Yen-Ting Lin\altaffilmark{1,2,3},
Jeremiah P.~Ostriker\altaffilmark{1},
and Christopher J.~Miller\altaffilmark{4}
}

\altaffiltext{1}{Department of Astrophysical Sciences, Princeton University,
Princeton, NJ 08544} 
\altaffiltext{2}{Departamento 
 de Astronom\'{i}a y Astrof\'{i}sica, Pontificia Universidad Cat\'{o}lica de Chile, Santiago, Chile}
\altaffiltext{3}{Current address: Institute for the Physics and Mathematics of the Universe, University of Tokyo, Kashiwa, Chiba, Japan; yen-ting.lin@ipmu.jp}
\altaffiltext{4}{Cerro Tololo Inter-American Observatory, La Serena, Chile}

\begin{abstract}

A novel statistic is proposed to examine the hypothesis that all cluster
galaxies are drawn from the same luminosity distribution (LD).  In such a
``statistical model'' of galaxy LD, the brightest cluster galaxies (BCGs) are
simply the statistical extreme of the galaxy population.  Using a large sample
of nearby clusters, we show that BCGs in high luminosity clusters (e.g.,
$L_{{\rm tot}}\gtrsim 4\times 10^{11} h_{70}^{-2} L_\odot$) are unlikely
(probability $\le 3\times 10^{-4}$) to be drawn from the LD defined by all red
cluster galaxies more luminous than $M_r= -20$. On the other hand, BCGs in
less luminous clusters are consistent with being the statistical extreme.
Applying our method to the second brightest galaxies, we show that they are
consistent with being the statistical extreme, which implies that the BCGs are
also distinct from non-BCG luminous, red, cluster galaxies.
We point out some issues with the interpretation of the classical tests
proposed by \citet[][TR]{tremaine77} that are designed to examine the
statistical nature of BCGs, investigate the robustness of both our statistical
test and those of TR against difficulties in photometry of galaxies of large
angular size, and discuss the implication of our findings on surveys that use
the luminous red galaxies to measure the baryon acoustic oscillation features
in the galaxy power spectrum.

\end{abstract}


\section{Introduction}
\label{sec:intro}

Over the past decade, the features in the cosmic microwave background angular power spectrum and galaxy clustering power spectrum due to the baryon acoustic oscillations (BAO) have emerged to be a premier standard ruler, and strong cosmological constraints have been derived using this technique \citep[e.g.,][]{spergel03,eisenstein05,cole05,padmanabhan07}.
In the galaxy BAO measurements, the tracer population of the large scale matter distribution often employed is the so-called luminous red galaxies (LRGs).
These are massive elliptical galaxies characterized by an old and passively evolving stellar population
\citep{eisenstein01}, and 
tend to be the
dominant, central galaxies in group or cluster scale dark matter halos [e.g., \citealt{zheng08};
throughout this paper we do not refer to the most luminous galaxy in clusters and groups differently, but simply use the term brightest cluster galaxies (BCGs)].

Recognizing the constraining power of BAO measurements,
an impressive array of on-going and planned cosmological experiments 
have adopted this method as the main survey component (e.g., BOSS\footnote{\url{http://www.sdss3.org/cosmology.php}}, WiggleZ\footnote{\url{http://wigglez.swin.edu.au/Welcome.html}}, HETDEX\footnote{\url{http://www.as.utexas.edu/hetdex/}}, ATLAS\footnote{\url{http://www.astro.dur.ac.uk/Cosmology/vstatlas/}}, 
ADEPT, WFMOS).
An important issue faced by these surveys is the control of systematics, including e.g., the corrections for the galaxy bias and the redshift distortion. In practical terms, it is necessary to know the statistical properties of the BAO tracer population to exquisite detail (such as the way they ``populate'' their host halos, and their luminosity function).
As the LRGs are used as the tracer in most of the aforementioned experiments, it is critical to 
test for the homogeneity of the LRG population. 
In this paper we
address two key questions:
{\it Are BCGs different from other cluster galaxies?
Are BCGs distinct from non-BCG, luminous, red galaxies in clusters and groups?}

Because of their brightness and uniformity in luminosity, the BCGs
have long been regarded as an ideal standard candle \citep[e.g.,][]{humason56,scott57}. 
Through the systematic investigations separately lead by Sandage and Gunn \citep[e.g.,][]{sandage72,sandage73a,sandage73b,gunn75,kristian78,hoessel80,schneider83a},
however, it was realized that various corrections [such as dependences of the BCG luminosity on the cluster richness class \citep{abell58} and Bautz-Morgan type \citep{bautz70}, and the effect of galactic cannibalism \citep{ostriker75}] have to be applied before any cosmological
implication from the BCG Hubble diagram can be extracted.

The small dispersion in BCG luminosity (e.g., $\sim 0.3$ mag as determined by \citealt{sandage72}) also
stimulated the discussion on their origin \citep[][to name a few]{scott57,peebles68,geller76,tremaine77,hausman78,geller83,merritt85,bhavsar85,lin04b,loh06,vonderlinden07,bernardi07,vale08}:
whether they are simply ``the brightest
of the bright'' or a different population from other cluster galaxies all together.
If the former hypothesis were true, the BCG luminosity distribution (LD) simply results from sampling
of an universal LD of all cluster galaxies, and its small dispersion in magnitude reflects the steepness of the bright end of the universal LD.
To test for such a ``statistical nature'' of the BCGs, \citet[][hereafter TR]{tremaine77} devised a couple of statistics based on the mean and dispersion of the magnitude difference between the first- and second-ranked galaxies ($\overline{\Delta}$ and $\sigma_{\Delta}$, respectively), and the dispersion of the BCG magnitude ($\sigma_1$).
The basic idea is that, if the BCGs are simply the statistical extreme of the cluster galaxy population, both $\sigma_1$ and $\sigma_{\Delta}$ would be greater or comparable to $\overline{\Delta}$. More specifically, under the assumptions that (1) 
the numbers of galaxies in non-overlapping magnitude intervals are independent random variables,
and (2) 
the LD drops to zero rapidly at the extreme bright end\footnote{This condition is solely to ensure the integrals behave well in the TR analysis. The ``extreme bright end'' can be at arbitrarily high luminosity (S.~Tremaine 2009, private communication). Other than these two assumptions, we note that the form of the LD can be very general for the TR tests to be applicable -- the LD does not even have to be continuous.},
the following conditions need to be satisfied: $T_1\equiv \sigma_1/\overline{\Delta}\ge 1$ and $T_2 \equiv \sigma_{\Delta}/\overline{\Delta}\gtrsim 0.82$
(see also \citealt{loh06}). 

Using LRGs to identify dense environments such as groups and clusters,
\citet{loh06} found a large magnitude gap ($\overline{\Delta}\sim 0.8$ mag) between the first- and second-ranked galaxies in $\sim 12,000$ LRG-selected groups and clusters (for which the majority of the LRGs are the BCGs). They concluded that the LRGs are inconsistent with being the statistical extreme of the universal galaxy LD, based on the tests proposed by TR.

One fundamental issue faced by studies of very luminous galaxies, such BCGs, concerns with their luminosity.
On the practical side, for nearby BCGs, measuring their total light is nontrivial. The majority of BCGs are giant elliptical or cD galaxies, 
whose surface brightness profile is typically flatter than that of normal elliptical galaxies, and can extend to tens or hundreds of kpc \citep{tonry87}. To measure the ``total'' magnitude requires careful subtraction of the sky background.
On the physical side, the cD envelope may well extend into the intracluster space, and it is sometimes difficult to separate the luminosity of the BCGs from that of the intracluster stars \citep[e.g.,][]{gonzalez05}.
In this study we will use data from the Sloan Digital Sky Survey (SDSS, \citealt{york00}), which is not deep enough to detect contributions from the intracluster stars. However, it is known that the current SDSS pipeline has difficulty handling photometry of galaxies with large angular extents, and therefore may seriously underestimate the luminosity of BCGs \citep[e.g.,][]{lauer07,vonderlinden07,sdssdr7}.
In such cases,
$\Delta$ will be biased low.
Although corrections to BCG photometry (if at all possible) may decrease the value of $T_1$, it may have the opposite effect on $T_2$, making the interpretation of the TR tests more difficult (see \S\ref{sec:photo}).

\begin{figure}
\epsscale{1.25}
\plotone{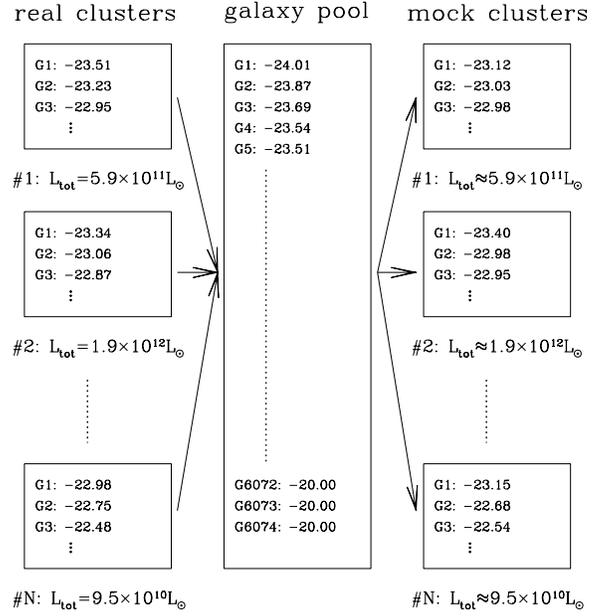}
\vspace{-7mm}
\caption{ 
An illustration of our method for obtaining the LD of all cluster galaxies, and for creating mock clusters. By combining member galaxies from all $N$ clusters, we get a ``galaxy pool'' that by definition should be representative of the LD. By drawing galaxies in random from the galaxy pool we then create $N$ clusters whose total luminosities are matched as close as possible to the real ones (see \S\ref{sec:mock} for details). Our galaxy selection is limited to $M_r=-20$.
}
\label{fig:pool}
\end{figure}

Here we propose a new way to examine the statistical nature of the BCGs that is less dependent on a robust measurement of the BCG luminosity and $\Delta$, 
and apply it to a large sample of nearby clusters.
Our method 
is both conceptually and operationally very simple: with a sample
of $N$ clusters, whose total galaxy luminosities are $L_{{\rm tot,}i}$, where $i=1,2,...N$, one can combine
all member galaxies to form a composite cluster. By drawing galaxies in random\footnote{In 
the process of creating the $N$ mock clusters, a galaxy can be picked more than once, or not selected at all.} from the composite cluster to create $N$ mock
clusters whose total luminosities are matched to the observed ones (see Fig.~\ref{fig:pool}), the most luminous mock galaxies
constitute the ``statistical'' BCG sample. One
can then examine the distribution
of the BCG luminosities ($L_{{\rm bcg}}$) for the observed and mock clusters.  In particular, we look for deviations from the statistical expectation from the correlation
between $L_{{\rm bcg}}$ and $L_{{\rm tot}}$.
As cluster luminosity is correlated with its mass (e.g., \citealt{lin04}), a comparison of the $L_{{\rm bcg}}$--$L_{{\rm tot}}$
correlation between the observed and mock clusters may provide some insights into possible cluster mass dependence (if any) in the BCG formation mechanism(s).

The great virtue of the TR tests is that they only rely on a minimal set of assumptions concerning the nature of the LD, and is immune to any cluster-to-cluster variation. Our proposed test, 
on the other hand, although relying on the universality of the LD, is independent of the shape of the LD.
In this sense the two approaches are highly complementary.

In \S\ref{sec:data} we describe our cluster sample and the galaxy data. 
We show the results of our tests in \S\ref{sec:thetest}, and present a comparison 
with the TR tests in \S\ref{sec:tr}. 
We apply our test to the second brightest cluster galaxies (hereafter G2s) in \S\ref{sec:g2}.
Extensive tests have been carried out in order to show the robustness of our results (\S\ref{sec:disc}).
We conclude by discussing the implications of our finding on the BCG formation scenarios and LRG-based BAO surveys in \S\ref{sec:disc_summary}. 

Throughout this paper we adopt a flat $\Lambda$CDM 
cosmological model where $\Omega_M=1-\Omega_\Lambda=0.3$
and $H_0=70 h_{70}\,{\rm km\,s}^{-1} {\rm Mpc}^{-1}$.

\section{Cluster Sample and Galaxy Data}
\label{sec:data}

Our clusters are drawn from an updated version of the C4 cluster catalog \citep{miller05}, which uses
data from the fifth data release of the SDSS \citep{sdssdr5}. Of the 2037 clusters in the sample, we restrict ourselves to 494 that lie within the redshift range $z=0.030-0.077$, with velocity dispersion $\sigma>200$ km/s, and contain at least 2 luminous galaxies (see below). While the lower limit in redshift helps to alleviate problems in photometric measurements of 
bright galaxies with large angular extents in SDSS, 
the upper limit is chosen to ensure that nearly all galaxies with $M_r\le-20$ have redshifts measured by SDSS (i.e., extinction-corrected petrosian magnitude $r_p\le 17.77$, among other selection criteria; \citealt{strauss02}).
Note that the characteristic magnitude for cluster red galaxies is $M_{*,r}=-21.70$ (see Table~\ref{tab:fits} below), based on our cluster sample.
In addition, as we will combine galaxies from all clusters to form a composite cluster, the small redshift
range chosen prevents any redshift evolution within our cluster galaxy sample.

Our first task is to 
assign cluster membership to galaxies in the SDSS {\em main} sample.
For every cluster, we include only {\it red} galaxies with redshift $|z-z_{cl}|\le 3\sigma/c$, with $r$-band absolute petrosian magnitude $M_r\le -20$, and are projected within $0.8h_{70}^{-1}$ Mpc of the cluster center (defined as the peak of bright galaxy density distribution; \citealt{miller05}). 
Here $z_{cl}$ and $c$ are the cluster redshift and speed of light, respectively.
A galaxy is considered red if the model color $u-r\ge 2.2$ \citep[e.g.,][]{strateva01}.
To calculate the absolute magnitudes, we use the $k$-correction based on the NYU value-added galaxy catalog \citep{blanton05}. In addition, a correction is made to convert the petrosian magnitudes into ``total'' magnitudes based on the surface brightness profile of the galaxies, following \citet{graham05}.

On average, about $94 \%$ of galaxies in the SDSS main sample have fibers assigned to them \citep{strauss02}. This fraction becomes smaller in crowded fields, such as the clusters (see \citealt{miller05} and below), due to the size of the fiber plugs. It is therefore critical to correct for such an incompleteness 
when trying to include all cluster galaxies using the SDSS main sample.
For every cluster we first calculate the mean $g-r$ and $r-i$ colors $\bar{c}_{gr}$ and $\bar{c}_{ri}$ of the spectroscopically confirmed member galaxies, then assume the galaxies that (1) satisfy the above selection criteria ($u-r\ge 2.2$, $r_p\le 17.7$, projected distance $\le 0.8 h_{70}^{-1}$ Mpc), (2) were targeted for spectroscopy but were not assigned a fiber, and (3) have $|g-r-\bar{c}_{gr}|\le 0.15$ and $|r-i-\bar{c}_{ri}|\le0.15$ as probable cluster members (hereafter referred to as the {\it photometric members}). 
We treat these galaxies exactly the same way as those with spectroscopic redshift measurements, and
derive the absolute magnitudes by assuming they are at the cluster redshift.

\begin{figure}
\epsscale{1.05}
\plotone{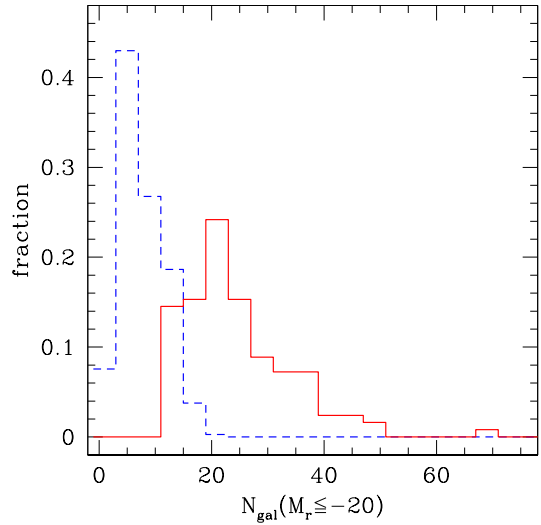}
\vspace{-2mm}
\caption{ 
The red/solid and blue/dashed histograms show the normalized distribution of number of galaxies for the high- and low-luminosity subsamples, respectively. More than 92\% of systems in the low-luminosity sample contain at least 5 red galaxies (more luminous than $M_r=-20$, located within $0.8 h_{70}^{-1}$ Mpc). 
}
\label{fig:rich}
\end{figure}

Combining the lists of the spectroscopic and photometric members, we record the absolute magnitudes
of the galaxies for each cluster.
The 5980 galaxies with $M_r\le -20$ from all 494 clusters form our composite cluster, or a ``galaxy pool'' that by definition has the statistical galaxy LD (see \S\ref{sec:intro}).
Among these galaxies,
871 are photometric members (including 97 BCGs).
In our analysis we will construct two cluster subsamples according to the total $r$-band luminosity $L_{{\rm tot}}$ from the member galaxies: a high-luminosity sample, consisting of 124 systems with $L_{{\rm tot}}>L_{\rm div}\equiv 3.7\times 10^{11} h_{70}^{-2} L_\odot$, and a low-luminosity sample with 370 systems whose $L_{{\rm tot}}\le L_{\rm div}$.
The cluster luminosity $L_{{\rm tot}}$ is the sum of luminosities from all red member galaxies (including the photometric members) more luminous than $M_r=-20$, projected within a radius of $0.8h_{70}^{-1}$ Mpc. 
As the SDSS main sample has complete spectroscopic coverage (modulo fiber collision) for $M_r\le -20$ for all our clusters,
no extrapolation in the luminosity function is needed to obtain $L_{{\rm tot}}$.
The division luminosity $L_{\rm div}$ is chosen so that each subsample contains roughly equal number of galaxies.

In Fig.~\ref{fig:rich} we show the distribution of number of galaxies for the two subsamples.
The histograms are normalized to the total number of clusters in the respective samples (red/solid: high-luminosity; blue/dashed: low-luminosity).
We see that $>92\%$ of low-luminosity systems have at least 5 red galaxies above our chosen magnitude limit. As we will show below, 
our results are not dependent on the faintest or most luminous systems.

\begin{figure}
\epsscale{1.05}
\plotone{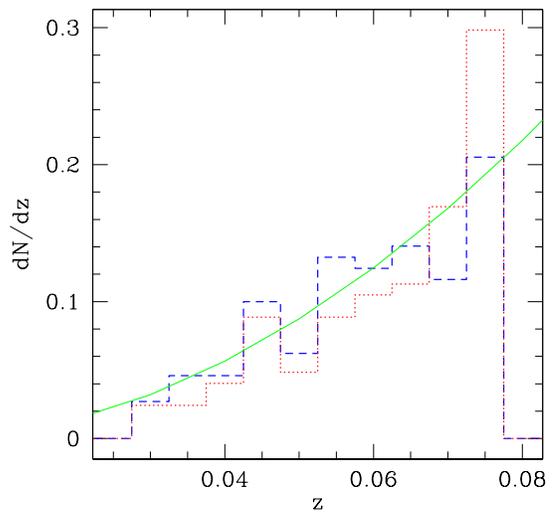}
\vspace{-2mm}
\caption{ 
The red/dotted and blue/dashed histograms show the normalized redshift distribution for the high- and low-luminosity subsamples, respectively. The two samples have similar distributions. 
For comparison, the green/solid curve shows the (arbitrarily) scaled differential comoving volume element $dV/dz$. As the redshift distributions roughly follow $dV/dz$, our samples are close to volume-limited.
}
\label{fig:zd}
\end{figure}

Fig.~\ref{fig:zd} presents the normalized redshift distribution $dN/dz$ of the two subsamples (red/dotted: high-luminosity; blue/dashed: low-luminosity). We find that the two samples have similar $dN/dz$: the median, mean, and $1\sigma$ scatter of the high-luminosity sample are
0.0662, 0.0624, and 0.0126, respectively. The same quantities for the low-luminosity sample are 0.0616, 0.0593, and 0.0126. The green/solid curve in the Figure is the differential comoving volume element $dV/dz$, scaled to the height of the distributions. The similarity between the redshift distributions and $dV/dz$ suggests that our samples are roughly volume-limited, and should be representative of the nearby groups and clusters.

\section{The $L_{{\lowercase{\rm bcg}}}$--$L_{{\lowercase {\rm tot}}}$ Correlation and the Statistical Nature of BCGs}
\label{sec:thetest}

Correlations between the luminosity of the BCGs and the properties of host clusters such as mass or total luminosity have been noted by previous studies \citep[e.g.,][]{lin04b,yang05,hansen07}. 
Here we utilize the BCG luminosity--cluster luminosity correlation ($L_{{\rm bcg}}$--$L_{{\rm tot}}$) to show that BCGs are indeed special -- at least in luminous/massive clusters.

In Fig.~\ref{fig:lbcgltot} (top panel) we show the observed $L_{{\rm bcg}}$--$L_{{\rm tot}}$ correlation
for the C4 clusters. 
The magenta squares in the top panel are the mean of BCG luminosity $\overline{L_{{\rm bcg,obs}}}$, as a function of $L_{{\rm tot}}$. The meaning of the cyan crosses will be described below.

\subsection{The Mock Clusters and BCGs}
\label{sec:mock}

To check for the statistical nature of BCGs, we proceed as follows. Based on the observed LD, we generate many realizations of the $L_{{\rm bcg}}$--$L_{{\rm tot}}$ correlation, from which we derive the mean $L_{{\rm bcg}}$--$L_{{\rm tot}}$ relation that is expected if the BCGs are drawn from the same LD as other cluster galaxies. We can then compare the observed mean $L_{{\rm bcg}}$--$L_{{\rm tot}}$ relation and determine if it is consistent with the mean from the mock data.

\begin{figure}
\epsscale{1.28}
\plotone{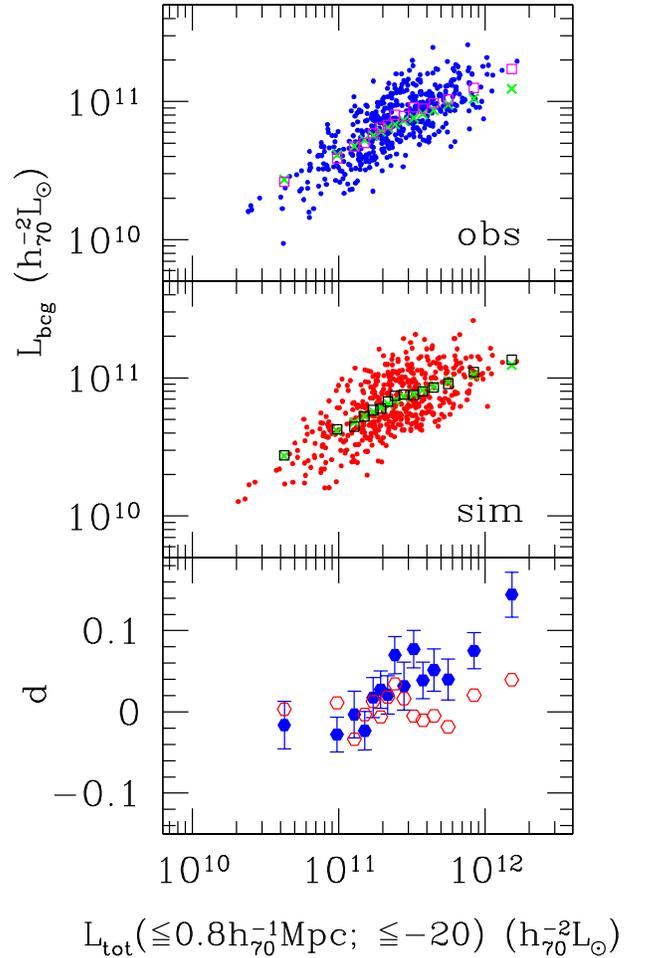}
\vspace{-6mm}
\caption{ 
  The $L_{{\rm bcg}}$--$L_{{\rm tot}}$ correlation for the real clusters (top panel) and one of the Monte Carlo realizations of mock cluster ensemble (middle panel). 
  The squares in the top and middle panel are the mean of the BCG luminosity for the real and the mock clusters ($\overline{L_{{\rm bcg,obs}}}$ and $\overline{L_{{\rm bcg,sim}}}$), respectively. Also shown in both panels as crosses are the mean BCG luminosity averaged over 200 Monte Carlo realizations, $\left< \overline{L_{{\rm bcg,sim}}} \right>$. In the bottom panel we show the distribution of $d_{{\rm obs}}= \log \overline{L_{{\rm bcg,obs}}} - \log \left< \overline{L_{{\rm bcg,sim}}}\right>$ (blue solid points) and $d_{{\rm sim}}= \log \overline{L_{{\rm bcg,sim}}} - \log \left< \overline{L_{{\rm bcg,sim}}}\right>$ (red open points). The real BCGs are systematically more luminous than the statistical ones in luminous clusters. The errorbars of $d_{{\rm obs}}$ are derived from 200 bootstrap resampling of the observed $L_{{\rm bcg}}$--$L_{{\rm tot}}$ correlation.
}
\label{fig:lbcgltot}
\end{figure}

We show in the middle panel of Fig.~\ref{fig:lbcgltot} one such realization. Basically, 
for each cluster in our sample, we create a corresponding mock cluster by 
randomly drawing galaxies from the galaxy pool, until the total luminosity of the mock cluster
matches that of the observed one (see also Fig.~\ref{fig:pool}).
Specifically, suppose the first $N$ galaxies give a total luminosity of $L_N<L_{{\rm tot}}$, and the next mock galaxy brings the total luminosity to $L_{N+1}>L_{{\rm tot}}$. We keep the $N+1$-th galaxy if $L_{N+1}-L_{{\rm tot}}<|L_N-L_{{\rm tot}}|$, otherwise we only use the first $N$ galaxies.
Because of this procedure, although the total luminosity for massive (luminous) clusters can be matched fairly well, for low luminosity clusters the difference in $L_{{\rm tot}}$ between the mock and real clusters can be large.

The black squares in the middle panels of Fig.~\ref{fig:lbcgltot} show the mean in BCG luminosity, $\overline{L_{{\rm bcg,sim}}}$ in this particular ensemble of simulated clusters. 
The cyan crosses in both the top and middle panels are the mean of $\overline{L_{{\rm bcg,sim}}}$ from 200 Monte Carlo realizations of the mock cluster ensemble, which we denote as $\left< \overline{L_{{\rm bcg,sim}}}\right>$. It is found that for luminous clusters (e.g., $L_{{\rm tot}}>3.7\times 10^{11} h_{70}^{-2} L_\odot$), the observed $\overline{L_{{\rm bcg,obs}}}$ is higher than $\left< \overline{L_{{\rm bcg,sim}}}\right>$ from mock clusters (e.g., compare the magenta squares with the cyan crosses in the top panel). For low luminosity clusters, the observed and mock values are comparable.

We list in Table~\ref{tab:def} the definition of various terms employed in this and the following subsections.

\begin{deluxetable*}{ll}

\tablecaption{Definition of Terms}

\tablehead{
\colhead{notation} & \colhead{meaning}
}

\startdata

$\overline{L_{{\rm bcg,obs}}}$ & mean BCG luminosity of the real clusters (in 15 $L_{{\rm tot}}$ bins)\\

$\overline{L_{{\rm bcg,sim}}}$ & mean BCG luminosity of one Monte Carlo ensemble (in 15 $L_{{\rm tot}}$ bins) \\

$\left< \overline{L_{{\rm bcg,sim}}}\right>$ & average of $\overline{L_{{\rm bcg,sim}}}$ over 200 Monte Carlo ensembles; this is the {\it statistical expectation}\\

$d_{{\rm obs}}$ & $\equiv  \log \overline{L_{{\rm bcg,obs}}} - \log \left< \overline{L_{{\rm bcg,sim}}}\right>$ \\

$d_{{\rm sim}}$ & $\equiv \log \overline{L_{{\rm bcg,sim}}} - \log \left< \overline{L_{{\rm bcg,sim}}}\right>$

\enddata

\label{tab:def}

\end{deluxetable*}

\subsection{Statistical Significance}
\label{sec:statsig}

By comparing the observed and mock $L_{{\rm bcg}}$--$L_{{\rm tot}}$ correlations we infer that BCGs in real clusters have systematically higher luminosities than the mock BCGs. 
One might easily imagine a systematic measurement error that systematically gives too large a luminosity for bright galaxies. But that would not produce the effect we find. One would need a systematic error that increased the brightness of BCGs but had a less dramatic effect on second brightest galaxies in other (more luminous) clusters of the same intrinsic luminosity.
Here we quantify the significance of the difference between the two populations.

\begin{figure}
\epsscale{1.15}
\plotone{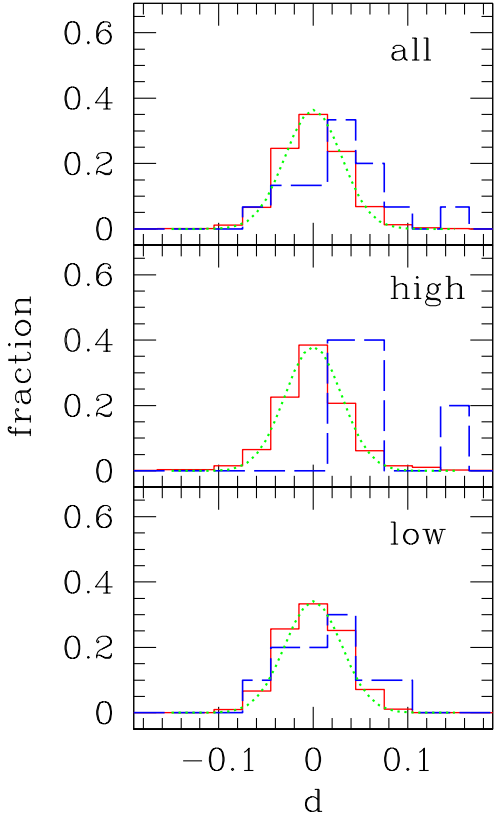}
\vspace{-1mm}
\caption{ 
  Distribution of the statistic $d$, defined as the difference between logarithms of the mean BCG luminosity $\overline{L_{{\rm bcg}}}$ and the mean of $\overline{L_{{\rm bcg,sim}}}$ from 200 Monte Carlo realizations of the mock cluster ensemble. The distribution of the observed $d_{{\rm obs}}$ is shown as the blue histograms, while that for the mock clusters is shown as the red ones. The top, middle, and bottom panels show the results for the whole cluster sample (494 systems), for high luminosity clusters ($L_{{\rm tot}}>L_{{\rm div}} = 3.7\times 10^{11} h_{70}^{-2} L_\odot$; 124 systems), and for low luminosity clusters ($L_{{\rm tot}}\le L_{{\rm div}}$; 370 systems), respectively. In each panel the green dotted curve is a Gaussian fit to the red histogram.
For the high luminosity clusters the difference between the real and mock data is significant at the 0.028\% level (see Table~\ref{tab:ours}).
}
\label{fig:sig}
\end{figure}

Using 200 realizations of the mock cluster ensemble (each containing 494 clusters), we calculate $\left< \overline{L_{{\rm bcg,sim}}}\right>$, as well as the difference between logarithms of the mean BCG luminosity $\overline{L_{{\rm bcg,sim}}}$ from individual realizations and $\left< \overline{L_{{\rm bcg,sim}}}\right>$, $d_{{\rm sim}}= \log \overline{L_{{\rm bcg,sim}}} - \log \left< \overline{L_{{\rm bcg,sim}}}\right>$. By dividing the cluster sample into 15 bins in $L_{{\rm tot}}$, each containing roughly equal number of clusters, for each mock cluster ensemble we have 15 evaluations of the statistic $d_{{\rm sim}}$. 
One such example is shown in the lower panel of Fig.~\ref{fig:lbcgltot} (open red symbols). We see that $d_{{\rm sim}}$ roughly scatters about zero.
We expect that for mock clusters, the distribution of $d_{{\rm sim}}$ should follow a Gaussian, which is shown as the blue histogram in the top panel of Fig.~\ref{fig:sig}. The green curve is a Gaussian fit to the histogram.

We can similarly calculate $d_{{\rm obs}} =  \log \overline{L_{{\rm bcg,obs}}} - \log \left< \overline{L_{{\rm bcg,sim}}}\right>$ for the real clusters. 
We show $d_{{\rm obs}}$ as a function of $L_{{\rm tot}}$ in the lower panel of Fig.~\ref{fig:lbcgltot} as the solid blue points; it is apparent that $d_{{\rm obs}}$ correlates positively with cluster luminosity. 
The errorbars of $d_{{\rm obs}}$ are derived from 200 bootstrap resampling of the observed $L_{{\rm bcg}}$--$L_{{\rm tot}}$ correlation.
The distribution of $d_{{\rm obs}}$ is further shown as the red dashed histogram in the top panel of Fig.~\ref{fig:sig}, and is clearly different from a Gaussian. A Kolmogorov-Smirnov (KS) test shows that the probability of $d_{{\rm obs}}$ to be drawn from the same distribution of $d_{{\rm sim}}$ is $\mathcal{P}=0.79\%$.

We examine in more detail the distribution of $d$ in high and low luminosity clusters in the other two panels of Fig.~\ref{fig:sig}. 
The distribution of $d$ for the high (low) luminosity subsample is shown in the middle (bottom) panel. As we note above, in high luminosity clusters the real BCGs are on average more luminous than the mock ones, resulting positive $d_{{\rm obs}}$, and there is only a 0.028\% probability that they are drawn from the same LD, based on a KS test. On the other hand, $\overline{L_{{\rm bcg,obs}}}$ in low luminosity clusters scatter about $\left< \overline{L_{{\rm bcg,sim}}}\right>$, and $d_{{\rm obs}}$ and $d_{{\rm sim}}$ have a much higher probability ($\mathcal{P}=54.5\%$) to be drawn from the same distribution. These results are recorded in Table~\ref{tab:ours} (the column under ``galaxy pool''; for meaning of the other columns, see \S\ref{sec:robust}).

\begin{deluxetable}{ccccc}

\tablecaption{Probability of BCGs being the Statistical Extreme}

\tablehead{
\colhead{} & \colhead{} & \multicolumn{3}{c}{method for creating mock clusters} \\ 
\cline{3-5} 
\colhead{sample} & \colhead{$N_{\rm cl}$\tablenotemark{a}} & \colhead{galaxy pool} & \colhead{Gaussian$+$Schechter\tablenotemark{b}} & \colhead{single Schechter\tablenotemark{b}}
}

\startdata
all    &  494 & \phn0.8\% & \phn0.5\% & \phn0.6\% \\
high & 124 & \phn\phn0.03\% & \phn\phn0.01\% & \phn\phn0.02\% \\
low   &  370 &     54.5\% &      39.0\% & 44.5\% 
\enddata

\tablenotetext{a}{number of clusters; the division of high and low luminosity subsamples is $3.7\times 10^{11} h_{70}^{-1} L_\odot$ (see \S\ref{sec:data}).}
\tablenotetext{b}{see \S\ref{sec:robust} for more details.}

\label{tab:ours}

\end{deluxetable}

\subsection{Tremaine-Richstone Tests}
\label{sec:tr}

Next we compare our results with those from the TR tests. 
Based on the whole cluster sample, we find that $\sigma_1=0.58$, $\overline{\Delta}=0.62$, and 
$\sigma_{\Delta}=0.49$, resulting in
$T_1=\sigma_1/\overline{\Delta} = 0.93$ and $T_2=\sigma_\Delta/\overline{\Delta}=0.78$.
Recall that if the BCGs are ``statistical'', we would have obtained $T_1\ge T_{{\rm 1,lim}} = 1$ and $T_2\gtrsim T_{{\rm 2,lim}} = 0.82$
(\S\ref{sec:intro}; but see below). Note that these lower limits are derived under rather general assumptions about the form of the LD (TR). For example, assuming the underlying LD follows the \citet{schechter76} form, the limits are $T_{{\rm 1,lim,sch}}\approx 1.16$ and $T_{{\rm 2,lim,sch}}\approx 0.88$.

Dividing the clusters into high and low luminosity subsamples (using the same division luminosity $L_{{}\rm div}$ as in \S\ref{sec:statsig}), 
we find
$(\sigma_1, \overline{\Delta}, \sigma_{\Delta}, T_1, T_2)=(0.39,0.55,0.43,0.70,0.79)$ and
$(0.54,0.65,0.50,0.84,0.77)$, respectively.
That $T_1$ for the whole sample is larger than $T_1$ for both high and low luminosity subsamples is due to the weak, positive correlation between $L_{{\rm bcg}}$ and $L_{{\rm tot}}$.

Both our test and the TR tests indicate that, using the whole cluster sample, the BCGs are not drawn from the same LD as other cluster galaxies. Separating the low luminosity clusters from the luminous ones, our test suggests that this conclusion is mainly driven by the BCGs in the luminous clusters. 
The $T_1$ statistic implies a larger 
deviation from the statistical expectation for BCGs in high luminosity clusters ($T_{{\rm 1,lim}}-T_1=0.30$), compared to their counterparts in less luminous clusters ($T_{{\rm 1,lim}}-T_1=0.16$).

\begin{figure*}
\epsscale{0.8}
\plotone{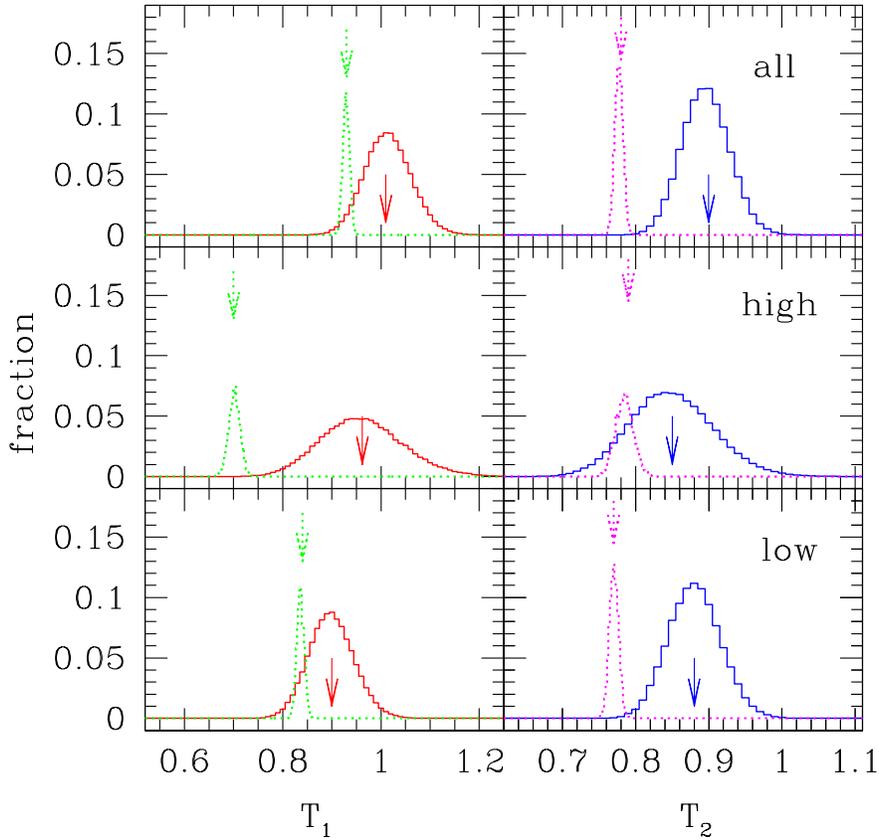}
\vspace{-5mm}
\caption{ 
Distribution of the $T_1$ and $T_2$ values for the real clusters (based on $5000$ bootstrap resampling) and mock clusters (derived from $10^5$ ensembles). The left panels show the distribution for $T_1$, while the right panels are for $T_2$. From top to bottom, we show the results for all the clusters, high and low luminosity subsamples, respectively. In each panel, the solid histogram is the distribution of the mock clusters (with the short solid arrow indicating the mean value), and the dotted histogram is that for the real clusters (with the mean indicated by the short dotted arrow).
The overlapping area under the solid and dotted histograms represents an estimate of the likelihood the observed value is consistent with the statistical expectations (see Table~\ref{tab:tr}).
Looking at the middle left panel (for $T_1$ values of high luminosity clusters), one sees that the dotted histogram is at the extreme of the solid red histogram, and would conclude that BCGs are unlikely to be statistical (based on the $T_1$ statistic). However, in the middle right panel (for $T_2$), the dotted histogram is well within the dotted blue histogram, suggesting the BCGs are statistical. The opposite situation happens when one looks at the bottom panels (for low luminosity clusters).
}
\label{fig:tr}
\end{figure*}

It is important to quantify how significant are the deviations of the $T_i$ ($i=1,2$) values we observe 
from the statistical limits.
Without assuming any particular form of the LD,
we evaluate the significance empirically, using the ``galaxy pool'' from all the member galaxies.
More specifically, we construct ensembles of mock clusters following the method described in \S\ref{sec:mock}; 
we can then derive the mean and dispersion of the two statistics from
the distributions of $T_1$ and $T_2$ based on the mock clusters (see Fig.~\ref{fig:tr}).
For all clusters in our sample, when the BCGs are generated statistically, the mean and standard deviation of the two statistics are $\overline{T_1}=1.01$, $\sigma_{T_1} = 0.05$, and $\overline{T_2}=0.90$, $\sigma_{T_2} = 0.03$. 
We find that about 3.5\% of the Monte
Carlo realizations have $T_1$ as low as the observed value (0.93), but only 0.007\% have $T_2$ as low
as 0.78, the observed value. We denote these fractions as $p_1 \equiv p_1(\le T_{{\rm 1,obs}})$ and $p_2\equiv p_2(\le T_{{\rm 2,obs}})$.
For high luminosity clusters, ($\overline{T_1}, \sigma_{T_1}, p_1, \overline{T_2}, \sigma_{T_2}, p_2)=(0.96,0.08,0.014\%,0.84,0.06,14\%)$.
For low luminosity ones, the values are $(0.90,0.05,8.8\%,0.88,0.04,0.07\%)$.

Additionally, we can estimate the distribution of the {\it observed} $T_i$ via bootstrap resampling. 
The overlapping area under the ``observed'' and the ``statistical'' distributions (the latter from the mock cluster ensembles, as mentioned above) then gives an estimate of the likelihood that the observed $T_i$ is consistent with the statistical expectation.
We denote the likelihood as $q_i \equiv q_i(T_{i,{\rm obs}} \cap T_{i,{\rm stat}})$, $i=1,2$.
For the whole cluster sample, we find $(q_1,q_2) = (8.8\%,1.4\%)$,
For high and low luminosity clusters, the results are $(0.35\%, 23.4\%)$ and $(14.9\%,0.64\%)$, respectively.

\begin{deluxetable*}{cccccccccc}

\tighten
\tabletypesize{\scriptsize}
\tablecaption{Tremaine-Richstone Statistics}
\tablewidth{0pt}

\tablehead{
\colhead{} & \multicolumn{3}{c}{all clusters} & \multicolumn{3}{c}{luminous clusters} &  \multicolumn{3}{c}{faint clusters} \\
\cline{2-4} \cline{5-7}  \cline{8-10} 
\colhead{} & \colhead{obs.\tablenotemark{a}} & \colhead{stat.\tablenotemark{b}} & \colhead{$(p,q)$\tablenotemark{c}} & \colhead{obs.\tablenotemark{a}}  & \colhead{stat.\tablenotemark{b}} & \colhead{$(p,q)$\tablenotemark{c}} & \colhead{obs.\tablenotemark{a}}  & \colhead{stat.\tablenotemark{b}} & \colhead{$(p,q)$\tablenotemark{c}}
}

\startdata

$T_1$ & (0.93,0.01) & $(1.01,0.05)$ & $(3.5,8.8)$ & (0.70,0.01) & $(0.96,0.08)$ & $(0.01,0.35)$ & (0.84,0.01) & $(0.90,0.05)$ & $(8.8,14.9)$\\
$T_2$ & (0.78,0.01) & $(0.90,0.03)$ & $(0.007,0.145)$ & (0.79,0.01) & $(0.85,0.06)$ & $(14,23)$ & (0.77,0.01) & $(0.88,0.04)$ & $(0.07,0.64)$

\enddata

\tablenotetext{a}{observed value and the standard deviation based on 5000 bootstrap resampling.}

\tablenotetext{b}{statistical expectation and the standard deviation based on $10^5$ Monte Carlo realizations.}

\tablenotetext{c}{$p$ is the proportion of the Monte Carlo realizations that give TR statistics as low as the observed value. $q$ is the area of the overlapping region under the observed and statistical distribution of the TR values (see Fig.~\ref{fig:tr}). Both $p$ and $q$ are in \%.}

\label{tab:tr}

\end{deluxetable*}

For clarity, these results are summarize in Table~\ref{tab:tr}.
We present the results for the whole sample, as well as the high and low luminosity subsamples.
For each (sub)sample, the numbers under ``obs.''~are the observed values, and the triplet of numbers
under ``stat.''~denotes the mean and standard deviation of the TR statistics based on $10^5$ ensembles of mock datasets, and $p_{i}$ ($i=1,2$).
Looking at $p_1$
for the high and low luminosity clusters, one would conclude that
BCGs in the faint clusters are much more likely to be statistical. One would draw the opposite conclusion if considering the $T_2$ statistic, however. 
It is therefore not clear if the TR tests give a consistent picture (e.g., dependence on $L_{{\rm tot}}$) for the degree of deviation of BCGs from the galaxy population in clusters.

Using a large group sample from SDSS, \citet{yang08} examined the gap statistics.
Although they did not explicitly calculate the values of $T_i$, using the results presented in their Fig.~7, we can roughly estimate that for clusters with virial mass $\log M \approx 14.8$, $(T_1,T_2) \sim (1.4,0.7)$, and for clusters with $\log M \approx 14.0$, $(T_1,T_2) \sim (0.6,0.8)$.
We see that the $T_i$ values have opposite trends with cluster mass: based on $T_1$, the BCGs in low mass systems are special; using $T_2$ one would be lead to the opposite conclusion.

In their study of LRG-selected groups and clusters, \citet{loh06} found that
$(\sigma_1, \overline{\Delta}, \sigma_{\Delta}, T_1, T_2)=(0.30,0.87,0.52,0.35,0.59)$
at $z\approx 0.12$. They also examined the richness dependence of
the TR statistics. For the richest systems ($75\%-100\%$ quartile in the richness distribution) they found $(T_1,T_2)=(0.75,0.71)$; for those with richness in the $25\%-50\%$ quartile (with about $2-3$ galaxies on the red sequence), $(T_1,T_2)=(0.27,0.34)$. 
The results of \citet{loh06} suggest 
a dependence of the TR statistics on richness that is opposite to our findings (if their richness estimator correlates well with the total luminosity).

The main cause for the differences seems to be in the magnitude gap, $\Delta$. Although for the richest systems, \citet{loh06} obtained $\overline{\Delta}\sim 0.55$, a value comparable to ours, their finding of $\overline{\Delta}\sim 1$ for the much poorer systems ($25\%-50\%$ quartile in richness) is much higher than our value.
Because \citet{loh06} looked for galaxy concentrations around LRGs 
using (primarily) photometric data, we suspect many of their low-richness systems may not be real, but simply
chance projections of galaxies with similarly red colors (c.f.~purity of the maxBCG algorithm; see \citealt{koester07}). 
On the other hand, 
we note that \citet{yang08} also found that $\overline{\Delta}\gtrsim 1$ for their low mass systems ($\log M \sim 13$).

These results seem to suggest that, when considered together, the two TR statistics may not unambiguously determine the statistical nature of BCGs with respect to cluster mass (or its proxy, such as total galaxy number or luminosity).

The disagreements among the different studies may be attributed to several factors (such as the way BCGs are selected, the way background contamination is treated, or even the group/cluster selection). A possible resolution would be to compare the selection of BCG and G2 on a cluster-by-cluster basis, for the clusters that are present in multiple catalogs.

\begin{figure}
\epsscale{1.15}
\plotone{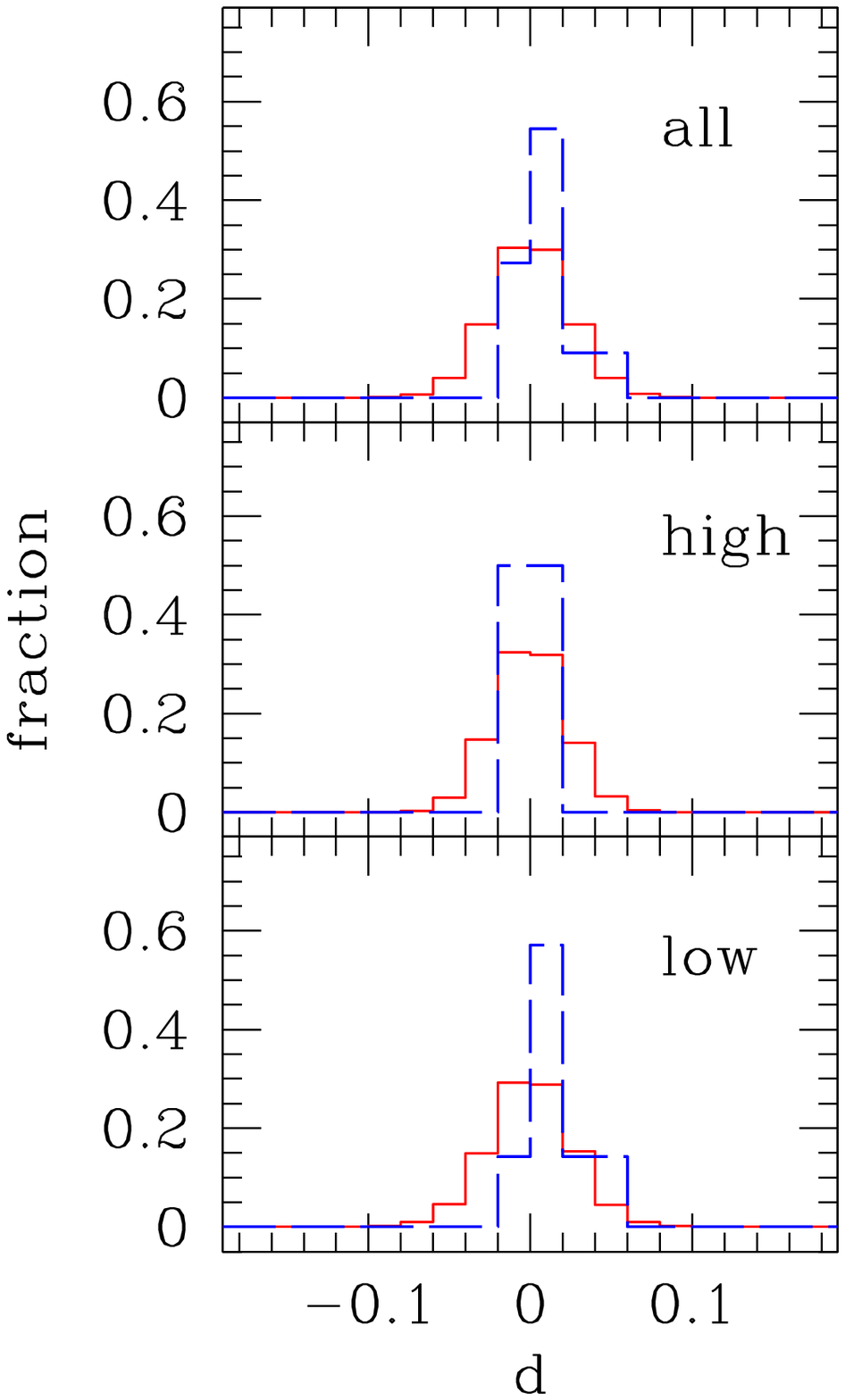}
\vspace{-1mm}
\caption{ 
Distribution of the statistic $d$ for G2. The distribution of the observed $d_{{\rm obs}}$ is shown as the blue/dashed histograms, while that for the mock clusters is shown as the red/solid ones. The bottom, middle, and top panels show the results for a low luminosity subsample ($1.6\times 10^{11} h_{70}^{-2} L_\odot \le L_{\rm tot} \le L_{\rm div}$; 234 systems), the high luminosity subsample ($L_{\rm tot}>L_{\rm div} = 3.7\times 10^{11} h_{70}^{-2} L_\odot$; 124 systems), and the combined sample (358 systems). In all samples considered, we find that G2s are consistent with the statistical expectation.
}
\label{fig:g2}
\end{figure}

\subsection{Are G2s Special?}
\label{sec:g2}

We have applied a new method to confirm that the BCGs are special. This approach can be used to check if G2s are statistical, that is, are their LD consistent with that of the overall cluster galaxy population. 
The results can then be used to answer the second question we set out to address:
{\it Are BCGs different from other luminous, red, cluster galaxies}?

We proceed in a similar fashion as in the previous sections. The main difference is that we are now comparing the observed $L_{{\rm g2}}$--$L_{{\rm tot}}$ correlation with the corresponding mean relation from mock clusters.
Requiring that mock clusters must have at least two members causes a small modification to the way mock clusters are constructed, which mainly affects the low luminosity clusters. Therefore we will set a lower luminosity limit at $L_{\rm lim} = 1.6\times 10^{11} h_{70}^{-2} L_\odot$, and define the low luminosity subsample (234 systems) with $L_{\rm lim}\le L_{\rm tot} \le L_{\rm div}$, where $L_{\rm div} = 3.7\times 10^{11} h_{70}^{-2} L_\odot$. The 124-cluster high luminosity subsample remains the same as before.

In Fig.~\ref{fig:g2} we show the distribution of the quantity $d$ for G2s, which is analogous to Fig.~\ref{fig:sig} for the BCGs. The bottom, middle, and top panels are the results for the low luminosity subsample, high luminosity subsample, and the combined sample (358 systems), respectively.
In all samples considered, we find that G2s are consistent with the statistical expectation.
KS tests suggest that $\mathcal{P}=26\%$, 55\%, and 24\% for all, high, and low luminosity clusters, respectively. 
Our results are not sensitive to the exact value of $L_{\rm lim}$.

This result immediately suggests that BCGs are distinct from G2s, and very likely, other luminous member galaxies.

This simple exercise also demonstrates one advantage of our method over the TR tests: while the latter by definition could not determine the statistical nature of G2s, our method can in principle be applied to even the third- or lower-ranked member galaxies.

\section{Systematics}
\label{sec:disc}

In this section we examine the robustness of our findings on various aspects in the analysis, including the selection of cluster red galaxies in SDSS, the way mock cluster and galaxy samples are constructed, and the issue of photometry of galaxies with large angular extent.

\subsection{Sensitivity on Galaxy and Cluster Sample Selection}
\label{sec:galsample}

We first check the sensitivity of our results on the galaxy selection criteria, by repeating our test using a much more stringent set of conditions to assign cluster memberships for red galaxies. In addition to the basic requirements ($u-r\ge 2.2$, $r_p\le 17.7$, projected within $0.8h_{70}^{-1}$~Mpc), for the {\it spectroscopic members}, we only include galaxies whose (1) redshifts satisfy $c|z-z_{cl}|\le v$, where $v\equiv {\rm min}(2\sigma, 1500~{\rm km/s})$, (2) $g-r$ and $r-i$ colors fall in the range $\bar{c}-1.5\sigma_{c}$ to  $\bar{c}+2.5\sigma_{c}$,
where $\bar{c}$ and $\sigma_c$ are the mean and dispersion of the $g-r$ and $r-i$ colors from the galaxies at redshifts broadly consistent with the cluster, (3) ``concentration parameter'' $c_m \equiv r_{90}/r_{50} \ge 2.6$, where $r_{90}$ and $r_{50}$ are radii that enclose 90\% and 50\% of the Petrosian flux, respectively; this last condition aims to select galaxies with early type morphology \citep{strateva01}.
As for the {\it photometric members}, 
along with the filters in the $r$-band flux, $u-r$ color, spatial distribution, and morphology,
we require that their $g-r$ and $r-i$ colors to lie between $\bar{c}-\sigma_{c}$ and $\bar{c}+1.5\sigma_{c}$, where $\bar{c}$ and $\sigma_c$ are now derived from the spectroscopic members.

These criteria reduce our cluster and galaxy sample; only 344 clusters containing 2819 galaxies are included.
We find that the probability $\mathcal{P}$ that $d_{{\rm obs}}$ and $d_{{\rm sim}}$ are drawn from the same distribution is 1.5\%, 0.15\%, and 67\% for the whole, high luminosity, and low luminosity (sub)samples.
These values confirm the results found in \S\ref{sec:statsig}.

We note in \S\ref{sec:data} that 20\% ($\approx 97/494$) of the BCGs are photometrically identified. Some of these may be foreground/background galaxies. 
To evaluate the effect of possible contamination due to these photometric BCGs,
we repeat our analysis using the 397 clusters whose BCGs are spectroscopically confirmed members (with the membership assignment criteria of \S\ref{sec:data}).
Based on the galaxy pool constructed from the 3661 galaxies in these clusters, we find that $\mathcal{P}=1.7\%$, 0.28\%, and 27\%, for the whole, high luminosity, and low luminosity (sub)samples.

Although the absolute values of $\mathcal{P}$ changes somewhat with respect to our nominal values recorded in Table~\ref{tab:ours}, the trend remains clear that the BCGs in high luminosity clusters are much less likely to be drawn from the same LD as other cluster members when compared to their counterparts in lower luminosity
clusters.

Let us comment next on one effect our cluster selection may have on the results. The requirement that the clusters need to host at least two galaxies with $M_r\le -20$ potentially excludes systems dominated by a single $\sim M_*$ galaxy, such as the ``fossil groups'' (which are defined to have $\Delta\ge 2$ and extended X-ray emissions). These systems are believed to evolve in isolation, with last major merger with other galactic systems being long enough in the past that a dominant central galaxy can result from the dynamic friction \citep{ponman94}.
The BCGs in these groups would then deviate significantly from the statistical extreme, which makes them distinct from the other BCGs in low luminosity groups included in our sample.
One might be concerned that the exclusion of fossil groups from
our cluster sample may have contributed to our conclusion that, for the lower luminosity
systems, there is no statistically significant deviation of BCGs from the statistical distribution of all cluster galaxies.
However, given the rarity of the fossils \citep[number density $\sim 2\times 10^{-6} h_{70}^{3}$~Mpc$^{-3}$; e.g.,][]{labarbera09,voevodkin09}, we expect there to be at most $\sim 20$ such systems in the volume we sample ($z=0.03-0.077$, $\approx 5700$ deg$^2$), and therefore our conclusion 
should be robust.

\subsection{Construction of Mock Clusters}
\label{sec:robust}

We first investigate whether the way galaxy pool is constructed affects our results. Instead of pouring galaxies from all clusters into one pool, we can create two pools separately for galaxies in high and low luminosity clusters. High and low luminosity mock clusters are then created from the respective pools. We find that $\mathcal{P}=0.98\%$ and 92\% for the high and low luminosity clusters, respectively. The trend is clear that BCGs in low luminosity clusters are far more likely to be the statistical extreme than their cousins in high luminosity systems.

\begin{deluxetable*}{ccccccccccc}

\tablecaption{Luminosity Distribution Parameters in $r$-band}

\tablehead{
\colhead{} & \multicolumn{3}{c}{single Schechter} & \multicolumn{7}{c}{Gaussian$+$Schechter\tablenotemark{a}}\\
\cline{2-4} \cline{5-11}
\colhead{sample} & \colhead{$\alpha$} & \colhead{$\phi_*$} & \colhead{$M_*$}  & \colhead{$\alpha$} & \colhead{$\phi_*$} & \colhead{$M_*$}  & \colhead{$M_{{\rm ch}}$} & \colhead{$\tilde{\phi}_{*}$} & \colhead{$\tilde{M}_{*}$} & \colhead{$\sigma_{M}^{2}$}
}

\startdata
all    & $-1.01$ & \phn8.53 & $-21.92$ & $-0.93$ & $10.25$ & $-21.70$ & $-22.40$ & 1.38 & $-22.25$ & 0.32 \\
high & $-1.17$ & 12.62 & $-22.13$ & $-0.99$ & $18.30$ & $-21.73$ & $-22.60$ & 1.90 & $-22.57$ & 0.25 \\
low   & $-0.93$ & \phn6.34 & $-21.78$ & $-0.73$ & \phn$8.92$ & $-21.28$ & $-22.05$ & 1.35 & $-21.98$ & 0.33
\enddata

\tablenotetext{a}{see Eq.~\ref{eq:gau_sch} for the definition of the parameters.}

\label{tab:fits}

\end{deluxetable*}

\begin{figure}
\epsscale{1.}
\plotone{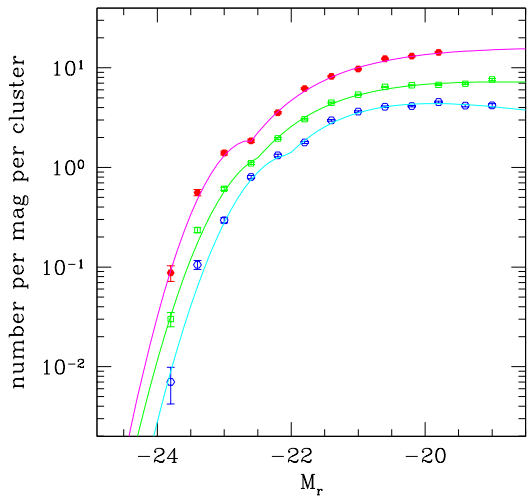}
\vspace{-4mm}
\caption{ 
  Composite LDs in the $r$-band for the 494 C4 clusters at $z=0.030-0.077$. The red, green, and blue symbols represent the LD for high luminosity clusters, the whole cluster sample, and low luminosity clusters, respectively. The curves are Gaussian$+$Schechter function fits to the LD (see Eq.~\ref{eq:gau_sch}); the best-fit parameters are given in Table~\ref{tab:fits}.
}
\label{fig:lfs}
\end{figure}

Secondly, we note that
instead of utilizing the galaxy pool, we can construct mock clusters using fits to the
observed LD of cluster galaxies.
The LDs for the whole cluster sample, and for the high and low luminosity subsamples, are measured by the method developed in \citet{lin04b}, to which we refer the reader for more details. 
The results are shown in Fig.~\ref{fig:lfs}. From top to bottom are the LDs for the high luminosity clusters, the whole sample, and the low luminosity clusters.

When BCGs are included, the \citeauthor{schechter76} function is not a good description of the LDs. 
Instead, the bright end may be more appropriately described by a log-normal distribution. 
Therefore, in fitting the LDs we consider both a bright end-truncated Schechter function plus a Gaussian, 
\begin{equation}
\label{eq:gau_sch}
  \phi(M) dM = \left\{ 
      \begin{array}{ll}
         \frac{\ln 10}{2.5} \phi_*  \mathcal{M}^{\alpha+1} \exp \left(- \mathcal{M} \right) dM    & \mbox{if $M\ge M_{{\rm ch}}$} \\
         \tilde{\phi}_{*}  \exp \left( -\frac{(M-\tilde{M}_{*})^2}{2\sigma_{M}^{2}} \right) dM                          & \mbox{otherwise}
      \end{array}
  \right.
\end{equation}
where $\mathcal{M}=10^{-0.4 (M-M_*)}$,
and a single Schechter function. The best-fit parameters are given in Table~\ref{tab:fits}.

Using the Gaussian$+$Schechter function fit to the whole sample as the LD, we find that for the whole, high, and low luminosity samples, 
$\mathcal{P}=0.47\%$, 0.013\%, and 39\%, respectively.
If using the Schechter function fit to the LD of the whole sample, 
$\mathcal{P}=0.57\%$, 0.015\%, and 44\% for the three (sub)samples, respectively.
These results are recorded in the third and fourth columns of Table~\ref{tab:ours}.

We thus conclude that our results are robust irrespective of the way the mock clusters are constructed. 
Another implication from this exercise is that our method does not rely on largely complete spectroscopic dataset like SDSS; as long as (1) one can determine the LD accurately (e.g., via a statistical background subtraction method; \citealt{lin04}), and (2) the cluster membership of BCGs can be reliably assigned, our test can be applied.

\subsection{Sensitivity on Photometry of BCGs}
\label{sec:photo}

We remark in \S\ref{sec:intro} concerning the difficulty in the measurement of the total light of the BCGs. Here we examine the effect of underestimation of BCG luminosity (and more generally, luminosity of galaxies with large angular extents) on both our test and the TR tests.
We address this issue by using two methods to obtain the best estimates of the true galaxy magnitudes for our galaxy sample.
The first one is a re-measurement of the galaxy photometry from the reduced SDSS images for every galaxy in our sample. The second one employs a statistical correction to the official SDSS photometry.

Simply put, the reason that the SDSS photometric pipeline 
has difficulties dealing with the photometry of large galaxies is because it is defaulted to use the ``local'' sky flux measured within regions of $256\times 256$ pixels ($1.7\arcmin \times 1.7\arcmin$) in the CCD frames. In the case that a large fraction of a given region is occupied by galaxies, the sky background level will be biased high, and in turn the galaxy magnitude will be lower than the true value.
A remedy is to use the ``global'' sky level measured in {\it frames} of $2048\times1498$ pixels ($13.5\arcmin\times9.8\arcmin$).
Following this line of reasoning, \citet[][]{vonderlinden07} devised a way to recover the true magnitude, given the crowdedness of a field, and the relative values of the local and global sky background.

As our first approach, we adopt the methodology of \citet{vonderlinden07} to apply corrections to the 
magnitudes of the galaxies 
and change the cluster luminosity accordingly, and then use the modified galaxy catalogs (and the galaxy pool) to conduct our test and the TR tests.
It is found that $\mathcal{P}=1.2\%$, 0.012\%, and 49\%, for the whole, high luminosity, and low luminosity cluster (sub)samples, respectively.
As for the TR tests, for the whole sample,
$(T_{{\rm 1}}, p_1,T_{{\rm 2}}, p_2)=(0.99,11.1\%,0.77,2\times10^{-4})$;
for the high and low luminosity subsamples,
$(T_{{\rm 1}}, p_1,T_{{\rm 2}}, p_2)=(0.68,5\times10^{-5},0.77,2.4\%)$ and
$(0.84,7.1\%,0.77,3\times10^{-4})$, respectively.

Using an independent photometric reduction software, \citet{hyde09} quantified the degree of underestimation of the magnitudes of large galaxies by the official SDSS pipeline. Their tests suggest that, for early type galaxies with effective radius $\approx 10\arcsec$, the mean magnitude deficit is about 0.25 mag, with a $68\%$ scatter about 0.04 mag.
They provided a fitting function that gives the mean deficit $\overline{\Delta m}$ as a function of the angular size $\theta$,
which we use as the basis of our second method for recovering the true galaxy magnitudes.
For every galaxy in our sample, we assume its magnitude was underestimated by the SDSS pipeline by
$\overline{\Delta m}(\theta)+\delta m$, where $\delta m$ is a Gaussian random variate with dispersion of 0.04 mag, and $\theta$ is determined from the de Vaucouleur fits to the galaxy surface brightness profile (see \citealt{hyde09} for details).

For the galaxy and cluster catalogs thus modified,
we find that $(T_{{\rm 1}}, T_{{\rm 2}})=(0.99,0.77)$ for the whole sample, and the probability to obtain $T_{1}$ and $T_2$ from the galaxy pool as low as the ``observed'' values is $11.3\%$ and $8\times 10^{-5}$, respectively. 
On the other hand, the probability that $d_{{\rm obs}}$ and $d_{{\rm sim}}$ are drawn from the same distribution is $\mathcal{P}=1.2\%$.
Breaking the sample into high and low luminosity subsamples, the results are
$(T_{{\rm 1}}, p_1,T_{{\rm 2}}, p_2)=(0.68,5\times10^{-5},0.78,2.2\%)$ for the luminous clusters, and
$(0.84,7.5\%,0.77,2\times 10^{-4})$ for the faint clusters.
From our test we find that BCGs in high and low luminosity clusters have probabilities of $\mathcal{P}=0.012\%$ and $\mathcal{P}=49\%$ to be statistical.

These results suggest that both our test and the TR tests are {\it insensitive} to the uncertainties in galaxy photometry due to sky subtraction.

It is also important to check if the low luminosity clusters are systematically at lower redshifts than the luminous ones. If this were true, the issue with sky subtraction of the SDSS pipeline may affect the photometry of BCGs in low luminosity clusters (hereafter LLCBCG) more strongly than that of BCGs in high luminosity clusters (hereafter HLCBCG), causing the photometry of LLCBCG to be underestimated more than the case for HLCBCG. This would in turn affect our finding that LLCBCG are consistent with being drawn from the same LD as other cluster galaxies (\S\ref{sec:statsig}).
We have shown in \S\ref{sec:data} that the high and low luminosity clusters have very similar redshift distribution, and therefore we do not think the photometry of LLCBCG and HLCBCG is treated differently.

\section{Discussions and Summary}
\label{sec:disc_summary}

\subsection{Implications on BCG Formation}
\label{sec:bcgform}

The main result of the present study is that BCGs as a whole have a LD that is distinct from that of the majority of red cluster galaxies (those with $M_r\le -20$).
This conclusion is primarily due to the high luminosities of HLCBCG, which has only 0.03\% of probability to be drawn from the LD of all galaxies. On the contrary, LLCBCG are more likely to be simply the statistical extreme of the LD of all galaxies.

With the help of mock clusters, and the elements that come into the calculation of TR tests, one gains some insight into the BCG formation process. For high luminosity clusters, we find that $(\sigma_1,\sigma_\Delta,\overline{\Delta})=(0.39,0.43,0.55)$ from the real data, while the corresponding values from the mock dataset are $(0.42,0.37,0.44)$. Therefore, although $\sigma_1$ in real and mock clusters are comparable, in real clusters $\overline{\Delta}$ is higher that that in mock ones (suggesting that real BCGs are on average 0.11 mag brighter than the statistical extrme). Physics of BCG formation drives the magnitude gap to be larger than the statistical expectation.

To explain the origin of HLCBCG, one therefore needs to invoke galactic mergers \citep[e.g.,][]{ostriker75,hausman78,lin04b,cooray05b,vale08}, which is naturally expected within the hierarchical structure formation paradigm \citep[e.g.,][]{dubinski98,delucia07}. However, the details of the mergers (e.g., major mergers between BCG and very luminous galaxies, minor mergers between BCG and $M_*$-type galaxies, or the ``galactic cannibalism''\footnote{We would like to point out that cannibalism proposed by Ostriker and co-workers is not binary mergers per se. It is based on the computable tendency of the most massive stellar systems to spiral to the centers of clusters via dynamical friction and to merge with other systems already resident at the center.}) remain to be understood. For example, while \citet{lin04b} suggested that major mergers are a viable route for forming HLCBCG, \citet{vale08} were more in favor of minor mergers.

In principle, the extent of late mergers can be constrained by the $L_{{\rm bcg}}$--$L_{{\rm tot}}$ correlation, or statistics related to the magnitude gap between BCG and second-ranked galaxy \citep[e.g.,][]{milosavlievic06,yang08}.
We have developed a simple merger model for cluster galaxies, and will present constraints on the importance of mergers based on the observations obtained in this paper in a future publication (Lin \& Ostriker 2009, in preparation).

After confirming that BCGs are indeed different from the bulk of cluster galaxy population, it is natural to ask if 
BCGs are different from other luminous/massive cluster galaxies (e.g., $M\lesssim M_*-1$).
It has long been known
that BCGs follow different scaling relations from other early type galaxies \citep[e.g.,][]{oegerle91,lauer07,desroches07,bernardi07}, in the sense that the BCGs are larger, less dense, and have lower velocity dispersion, compared to other early type galaxies of the same luminosity.
However, at the present it is not clear if the structure of the BCGs (in terms of Sersic fits) is indeed different from non-BCG early type galaxies of comparable color and stellar mass (see \citealt{guo09}).

We address this question somewhat indirectly; in \S\ref{sec:g2} it is shown that the LD of G2s in high luminosity clusters are consistent with that of the whole cluster galaxy population. (Our method could not be robustly applied to examine the statistical nature of G2s in low luminosity clusters, unfortunately.) Although we do not check the third-ranked brightest members, we believe a similar result will emerge for them.
Effectively, we suggest that among the most luminous, red, cluster members, only BCGs in high luminosity clusters show significant deviations from the statistical extreme, and therefore these galaxies are a distinct population.

Our results suggest both BCGs in low luminosity/mass clusters and G2s in luminous/massive clusters are the statistical extreme of the galaxy LD. As the latter are very likely BCGs (in clusters or groups that merge with the current host clusters) themselves at earlier stages of their lives, this finding seems to give a self-consistent
picture of the luminous galaxy evolution within the hierarchical cluster evolution scenario.

\subsection{Cosmological Implications}
\label{sec:implication}

The results and reasoning presented in \S\ref{sec:g2} \& \S\ref{sec:bcgform} suggest
that LRGs that are BCGs in high luminosity (massive) clusters are different from other LRGs (e.g., those that are LLCBCG and G2s or lower-ranked galaxies in high luminosity clusters), and call for very careful modeling and selection of the LRGs.

Suppose the finding that HLCBCG and LLCBCG 
are two distinct populations continues to hold towards higher redshifts. In a flux-limited survey, the LRGs at higher redshifts would be intrinsically more luminous than those at lower-$z$.
Therefore, a larger fraction of LRGs at higher-$z$ would be composed of HLCBCG, when compared to LRGs at lower-$z$.
When accounting for the Malmquist bias present in the LRG sample, one cannot simply assume that BCGs follow the same LD as LRGs as a whole, otherwise the inferred luminosity would be lower than the true value. 
This potential systematic effect would be larger towards higher-$z$.

It is thus important to take into account the difference in the LD of HLCBCG and LLCBCG when creating mock LRG catalogs for BAO surveys.
Unfortunately the BCGs in the present study are at lower redshifts ($z<0.1$) even compared to SDSS and SDSS-II LRG studies \citep[at $z\sim 0.3$; e.g.,][]{eisenstein05}, and therefore the LDs we measure (see the Gaussian fits in Table~\ref{tab:fits}) are not readily applicable.
Accordingly our study strongly motivates for a systematic investigation on the $L_{{\rm bcg}}$--$L_{{\rm tot}}$ correlation at $z>0.3$, either through direct observations (similar to the methodology used in the present paper, utilizing spectroscopic data from e.g., BOSS or GAMA surveys
), through the halo occupation distribution analysis \citep[e.g.,][]{zheng08}, or the non-parametric method of \citet{vale06}.

\subsection{Summary}
\label{sec:summary}

The question of whether the luminosity distribution of BCGs is drawn from the same LD as other cluster galaxies has been extensively discussed over the last four decades. Here we propose a simple new test to examine the statistical nature of the BCGs, and 
to supplement
the classical tests proposed by \citet[][TR]{tremaine77}.

Our basic idea is to shuffle the cluster galaxies and see how likely the observed LD of BCGs can be reproduced. The procedure is sketched in Fig.~\ref{fig:pool}.
Using a sample of 494 clusters from the C4 catalog at $z=0.03-0.077$, we combine all member galaxies  with $M_r\le-20$ to form a big pool of galaxies (5980 in total). We then randomly pick up galaxies from the pool to create 494 mock clusters, whose total luminosities $L_{{\rm tot}}$ are matched to that of the real clusters as close as possible.
This way we can compare the correlation between the BCG luminosity $L_{{\rm bcg}}$ and $L_{{\rm tot}}$ of real and mock clusters (see Fig.~\ref{fig:lbcgltot}). Of course, there are more than one way to create an ensemble of mock clusters, so we repeat this Monte Carlo process to generate 200 mock ensembles.

The averaged BCG luminosity from all 200 realizations of mock datasets, $\left< \overline{L_{{\rm bcg,sim}}}\right>$, gives the expected value when BCGs are statistical, as a function of cluster luminosity. The differences between the logarithms of mean mock BCG luminosity ($\overline{L_{{\rm bcg,sim}}}$) from each ensemble and $\left< \overline{L_{{\rm bcg,sim}}}\right>$ represent the degree of scatter expected if BCGs are stochastically selected from a global LD. In Fig.~\ref{fig:sig} we compare the distribution of 
$d_{{\rm sim}}= \log \overline{L_{{\rm bcg,sim}}} - \log \left< \overline{L_{{\rm bcg,sim}}}\right>$ with 
that of
$d_{{\rm obs}}= \log \overline{L_{{\rm bcg,obs}}} - \log \left< \overline{L_{{\rm bcg,sim}}}\right>$, and find that the real BCGs are more luminous than the statistical expectation; there is 0.8\% of probability for 
$d_{{\rm obs}}$ to be drawn from the same distribution of $d_{{\rm sim}}$ (see Table~\ref{tab:ours}).
Separating the clusters into high and low luminosity subsamples (with a division luminosity of $L_{{\rm div}} = 3.7\times 10^{11} h_{70}^{-2} L_\odot$),
we conclude that the difference is mainly coming from BCGs in high luminosity clusters (only 0.03\% chance to be statistical). The luminosities of BCGs in low luminosity clusters are roughly consistent with the global LD.

We extend the analysis to discuss the statistical nature of the second brightest galaxies (G2s), and find strong evidence that (in high luminosity clusters), G2s have a LD similar to that of the whole cluster galaxy population.

We also apply the tests proposed by TR to our cluster sample, and record the results in Table~\ref{tab:tr}.
The two statistics ($T_1$ and $T_2$) both suggest a small probability for BCGs to be statistical, consistent with our findings. However, examining the TR statistics for the high and low luminosity clusters reveals confusing trends. According to the $T_1$ statistic, BCGs in the high luminosity clusters are much less likely to be statistical than their counterparts in the low luminosity clusters. However, the $T_2$ statistic suggests the opposite conclusion.

Our results confirm previous findings that BCGs in high luminosity clusters are a distinct population from other cluster galaxies and tentatively supports the physical mechanism of cannibalism \citep{ostriker75} which is essentially dynamical friction. 
We also suggest that BCGs are distinct from other non-BCG LRGs.
As the majority of LRGs are the brightest member in groups and clusters, such an effect should be taken into account in selecting a homogeneous sample of LRGs for on-going and future BAO surveys.

\acknowledgments

We are grateful to Scott Tremaine and Bob Nichol for insightful comments on the manuscript.
YTL thanks Michael Strauss, Antonio Vale, and Jim Gunn for helpful discussions,
and IH for constant encouragement.
YTL acknowledges supports from the Princeton-Cat\'{o}lica Fellowship, 
NSF PIRE grant OISE-0530095, FONDAP-Andes,
and the World Premier International Research Center Initiative, MEXT, Japan.

Funding for the SDSS and SDSS-II has been provided
by the Alfred P. Sloan Foundation, the Participating Institutions, the National
Science Foundation, the U.S. Department of Energy, the National Aeronautics and
Space Administration, the Japanese Monbukagakusho, and the Max Planck Society,
and the Higher Education Funding Council for England. The SDSS Web site is
http://www.sdss.org/.

The SDSS is managed by the Astrophysical Research Consortium for the
Participating Institutions. The Participating Institutions are the American
Museum of Natural History, Astrophysical Institute Potsdam, University of
Basel, University of Cambridge, Case Western Reserve University, The University
of Chicago, Drexel University, Fermilab, the Institute for Advanced Study, the
Japan Participation Group, The Johns Hopkins University, the Joint Institute
for Nuclear Astrophysics, the Kavli Institute for Particle Astrophysics and
Cosmology, the Korean Scientist Group, the Chinese Academy of Sciences, Los Alamos National Laboratory, the Max-Planck-Institute for
Astronomy, the Max-Planck-Institute for Astrophysics, New Mexico
State University, Ohio State University, University of Pittsburgh, University
of Portsmouth, Princeton University, the United States Naval Observatory, and
the University of Washington.

\bibliographystyle{apj}

\begin{thebibliography}{50}
\expandafter\ifx\csname natexlab\endcsname\relax\def\natexlab#1{#1}\fi


\bibitem[{{Abazajian} {et~al.}(2008)}]{sdssdr7}
{Abazajian}, K., et al.~2008, \apjs, submitted (arXiv:0812.0649)


\bibitem[{{Abell}(1958)}]{abell58}
{Abell}, G.~O. 1958, \apjs, 3, 211

\bibitem[Adelman-McCarthy et al.(2007)]{sdssdr5} 
Adelman-McCarthy, J.~K., et al.\ 2007, \apjs, 172, 634 

\bibitem[{{Bautz} \& {Morgan}(1970)}]{bautz70}
{Bautz}, L.~P. \& {Morgan}, W.~W. 1970, \apjl, 162, L149

\bibitem[{{Bernardi} {et~al.}(2007){Bernardi}, {Hyde}, {Sheth}, {Miller}, \&
  {Nichol}}]{bernardi07}
{Bernardi}, M., {Hyde}, J.~B., {Sheth}, R.~K., {Miller}, C.~J., \& {Nichol},
  R.~C. 2007, \aj, 133, 1741

\bibitem[{{Bhavsar} \& {Barrow}(1985)}]{bhavsar85}
{Bhavsar}, S.~P. \& {Barrow}, J.~D. 1985, \mnras, 213, 857

\bibitem[{{Blanton} {et~al.}(2005){Blanton}, {Schlegel}, {Strauss},
  {Brinkmann}, {Finkbeiner}, {Fukugita}, {Gunn}, {Hogg}, {Ivezi{\'c}}, {Knapp},
  {Lupton}, {Munn}, {Schneider}, {Tegmark}, \& {Zehavi}}]{blanton05}
{Blanton}, M.~R., {Schlegel}, D.~J., {Strauss}, M.~A., {Brinkmann}, J.,
  {Finkbeiner}, D., {Fukugita}, M., {Gunn}, J.~E., {Hogg}, D.~W., {Ivezi{\'c}},
  {\v Z}., {Knapp}, G.~R., {Lupton}, R.~H., {Munn}, J.~A., {Schneider}, D.~P.,
  {Tegmark}, M., \& {Zehavi}, I. 2005, \aj, 129, 2562


  
\bibitem[Cole et al.(2005)]{cole05} Cole, S., et al.\ 2005, 
\mnras, 362, 505 
  

\bibitem[{{Cooray} \& {Milosavljevi{\'c}}(2005)}]{cooray05b}
{Cooray}, A. \& {Milosavljevi{\'c}}, M. 2005, \apjl, 627, L85

\bibitem[{{De Lucia} \& {Blaizot}(2007)}]{delucia07}
{De Lucia}, G. \& {Blaizot}, J. 2007, \mnras, 375, 2

\bibitem[Desroches et al.(2007)]{desroches07} Desroches, L.-B., 
Quataert, E., Ma, C.-P., \& West, A.~A.\ 2007, \mnras, 377, 402 



\bibitem[{{Dubinski}(1998)}]{dubinski98}
{Dubinski}, J. 1998, \apj, 502, 141



\bibitem[Eisenstein et al.(2001)]{eisenstein01} Eisenstein, D.~J., 
et al.\ 2001, \aj, 122, 2267 


  
  \bibitem[Eisenstein et al.(2005)]{eisenstein05} Eisenstein, D.~J., 
et al.\ 2005, \apj, 633, 560 



\bibitem[{{Geller} \& {Peebles}(1976)}]{geller76}
{Geller}, M.~J. \& {Peebles}, P.~J.~E. 1976, \apj, 206, 939

\bibitem[{{Geller} \& {Postman}(1983)}]{geller83}
{Geller}, M.~J. \& {Postman}, M. 1983, \apj, 274, 31

\bibitem[{{Gonzalez} {et~al.}(2005){Gonzalez}, {Zabludoff}, \&
  {Zaritsky}}]{gonzalez05}
{Gonzalez}, A.~H., {Zabludoff}, A.~I., \& {Zaritsky}, D. 2005, \apj, 618, 195

\bibitem[{{Graham} {et~al.}(2005){Graham}, {Driver}, {Petrosian}, {Conselice},
  {Bershady}, {Crawford}, \& {Goto}}]{graham05}
{Graham}, A.~W., {Driver}, S.~P., {Petrosian}, V., {Conselice}, C.~J.,
  {Bershady}, M.~A., {Crawford}, S.~M., \& {Goto}, T. 2005, \aj, 130, 1535

\bibitem[{{Gunn} \& {Oke}(1975)}]{gunn75}
{Gunn}, J.~E. \& {Oke}, J.~B. 1975, \apj, 195, 255

\bibitem[Guo et al.(2009)]{guo09} Guo, Y., et al.\ 2009, \mnras, submitted (arXiv:0901.1150) 



\bibitem[{{Hansen} {et~al.}(2007){Hansen}, {Sheldon}, {Wechsler}, \&
  {Koester}}]{hansen07}
{Hansen}, S.~M., {Sheldon}, E.~S., {Wechsler}, R.~H., \& {Koester}, B.~P. 2007,
  \apj, submitted (arXiv:0710.3780)

\bibitem[{{Hausman} \& {Ostriker}(1978)}]{hausman78}
{Hausman}, M.~A. \& {Ostriker}, J.~P. 1978, \apj, 224, 320

\bibitem[{{Hoessel} {et~al.}(1980){Hoessel}, {Gunn}, \& {Thuan}}]{hoessel80}
{Hoessel}, J.~G., {Gunn}, J.~E., \& {Thuan}, T.~X. 1980, \apj, 241, 486

\bibitem[{{Humason} {et~al.}(1956){Humason}, {Mayall}, \&
  {Sandage}}]{humason56}
{Humason}, M.~L., {Mayall}, N.~U., \& {Sandage}, A.~R. 1956, \aj, 61, 97


\bibitem[Hyde \& Bernardi(2009)]{hyde09} Hyde, J.~B., \& Bernardi, M.\ 2009, \mnras, accepted (arXiv:0810.4922) 


\bibitem[Koester et al.(2007)]{koester07} Koester, B.~P., et al.\ 
2007, \apj, 660, 221 





\bibitem[{{Kristian} {et~al.}(1978){Kristian}, {Sandage}, \&
  {Westphal}}]{kristian78}
{Kristian}, J., {Sandage}, A., \& {Westphal}, J.~A. 1978, \apj, 221, 383


\bibitem[La Barbera et al.(2009)]{labarbera09} La Barbera, F., de 
Carvalho, R.~R., de la Rosa, I.~G., Sorrentino, G., Gal, R.~R., 
\& Kohl-Moreira, J.~L.\ 2009, \aj, 137, 3942 


\bibitem[{{Lauer} {et~al.}(2007){Lauer}, {Faber}, {Richstone}, {Gebhardt},
  {Tremaine}, {Postman}, {Dressler}, {Aller}, {Filippenko}, {Green}, {Ho},
  {Kormendy}, {Magorrian}, \& {Pinkney}}]{lauer07}
{Lauer}, T.~R., {Faber}, S.~M., {Richstone}, D., {Gebhardt}, K., {Tremaine},
  S., {Postman}, M., {Dressler}, A., {Aller}, M.~C., {Filippenko}, A.~V.,
  {Green}, R., {Ho}, L.~C., {Kormendy}, J., {Magorrian}, J., \& {Pinkney}, J.
  2007, \apj, 662, 808

\bibitem[{{Lin} \& {Mohr}(2004)}]{lin04b}
{Lin}, Y.-T. \& {Mohr}, J.~J. 2004, \apj, 617, 879

\bibitem[{{Lin} {et~al.}(2004){Lin}, {Mohr}, \& {Stanford}}]{lin04}
{Lin}, Y.-T., {Mohr}, J.~J., \& {Stanford}, S.~A. 2004, \apj, 610, 745

\bibitem[{{Loh} \& {Strauss}(2006)}]{loh06}
{Loh}, Y.-S. \& {Strauss}, M.~A. 2006, \mnras, 366, 373

\bibitem[{{Merritt}(1985)}]{merritt85}
{Merritt}, D. 1985, \apj, 289, 18

\bibitem[{{Miller} {et~al.}(2005){Miller}, {Nichol}, {Reichart}, {Wechsler},
  {Evrard}, {Annis}, {McKay}, {Bahcall}, {Bernardi}, {Boehringer}, {Connolly},
  {Goto}, {Kniazev}, {Lamb}, {Postman}, {Schneider}, {Sheth}, \&
  {Voges}}]{miller05}
{Miller}, C.~J., {Nichol}, R.~C., {Reichart}, D., {Wechsler}, R.~H., {Evrard},
  A.~E., {Annis}, J., {McKay}, T.~A., {Bahcall}, N.~A., {Bernardi}, M.,
  {Boehringer}, H., {Connolly}, A.~J., {Goto}, T., {Kniazev}, A., {Lamb}, D.,
  {Postman}, M., {Schneider}, D.~P., {Sheth}, R.~K., \& {Voges}, W. 2005, \aj,
  130, 968

\bibitem[{{Milosavljevi{\'c}} {et~al.}(2006){Milosavljevi{\'c}}, {Miller},
  {Furlanetto}, \& {Cooray}}]{milosavlievic06}
{Milosavljevi{\'c}}, M., {Miller}, C.~J., {Furlanetto}, S.~R., \& {Cooray}, A.
  2006, \apjl, 637, L9

\bibitem[Oegerle \& Hoessel(1991)]{oegerle91} Oegerle, W.~R., \& Hoessel, J.~G.\ 1991, \apj, 375, 15 


\bibitem[{{Ostriker} \& {Tremaine}(1975)}]{ostriker75}
{Ostriker}, J.~P. \& {Tremaine}, S.~D. 1975, \apjl, 202, L113



\bibitem[Padmanabhan et al.(2007)]{padmanabhan07} Padmanabhan, N., et 
al.\ 2007, \mnras, 378, 852 




\bibitem[{{Peebles}(1968)}]{peebles68}
{Peebles}, P.~J.~E. 1968, \apj, 153, 13

\bibitem[Ponman et al.(1994)]{ponman94} Ponman, T.~J., Allan, 
D.~J., Jones, L.~R., Merrifield, M., McHardy, I.~M., Lehto, H.~J., 
\& Luppino, G.~A.\ 1994, \nat, 369, 462 




\bibitem[{{Sandage}(1972)}]{sandage72}
{Sandage}, A. 1972, \apj, 178, 1

\bibitem[{{Sandage}(1973)}]{sandage73a}
---. 1973, \apj, 183, 731

\bibitem[{{Sandage} \& {Hardy}(1973)}]{sandage73b}
{Sandage}, A. \& {Hardy}, E. 1973, \apj, 183, 743

\bibitem[{{Schechter}(1976)}]{schechter76}
{Schechter}, P. 1976, \apj, 203, 297

\bibitem[{{Schneider} {et~al.}(1983){Schneider}, {Gunn}, \&
  {Hoessel}}]{schneider83a}
{Schneider}, D.~P., {Gunn}, J.~E., \& {Hoessel}, J.~G. 1983, \apj, 264, 337

\bibitem[{{Scott}(1957)}]{scott57}
{Scott}, E.~L. 1957, \aj, 62, 248

\bibitem[{{Spergel} {et~al.}(2003){Spergel}, {Verde}, {Peiris}, {Komatsu},
  {Nolta}, {Bennett}, {Halpern}, {Hinshaw}, {Jarosik}, {Kogut}, {Limon},
  {Meyer}, {Page}, {Tucker}, {Weiland}, {Wollack}, \& {Wright}}]{spergel03}
{Spergel}, D.~N., {Verde}, L., {Peiris}, H.~V., {Komatsu}, E., {Nolta}, M.~R.,
  {Bennett}, C.~L., {Halpern}, M., {Hinshaw}, G., {Jarosik}, N., {Kogut}, A.,
  {Limon}, M., {Meyer}, S.~S., {Page}, L., {Tucker}, G.~S., {Weiland}, J.~L.,
  {Wollack}, E., \& {Wright}, E.~L. 2003, \apjs, 148, 175


  
  \bibitem[Strateva et al.(2001)]{strateva01} Strateva, I., et al.\ 
2001, \aj, 122, 1861 




  
  
\bibitem[Strauss et al.(2002)]{strauss02} Strauss, M.~A., et al.\ 
2002, \aj, 124, 1810 




\bibitem[{{Tonry}(1987)}]{tonry87}
{Tonry}, J.~L. 1987, in IAU Symp. 127: Structure and Dynamics of Elliptical
  Galaxies, 89--96

\bibitem[{{Tremaine} \& {Richstone}(1977)}]{tremaine77}
{Tremaine}, S.~D. \& {Richstone}, D.~O. 1977, \apj, 212, 311

\bibitem[{{Vale} \& {Ostriker}(2006)}]{vale06}
{Vale}, A. \& {Ostriker}, J.~P. 2006, \mnras, 371, 1173

\bibitem[{{Vale} \& {Ostriker}(2008)}]{vale08}
---. 2008, \mnras, 383, 355


\bibitem[Voevodkin et al.(2009)]{voevodkin09} Voevodkin, A., 
Borozdin, K., Heitmann, K., Habib, S., Vikhlinin, A., Mescheryakov, A., 
\& Hornstrup, A.\ 2009, \apj, submitted (arXiv:0902.0619)



\bibitem[{{von der Linden} {et~al.}(2007){von der Linden}, {Best}, {Kauffmann},
  \& {White}}]{vonderlinden07}
{von der Linden}, A., {Best}, P.~N., {Kauffmann}, G., \& {White}, S.~D.~M.
  2007, \mnras, 379, 867

\bibitem[{{Yang} {et~al.}(2005){Yang}, {Mo}, {Jing}, \& {van den
  Bosch}}]{yang05}
{Yang}, X., {Mo}, H.~J., {Jing}, Y.~P., \& {van den Bosch}, F.~C. 2005, \mnras,
  358, 217

\bibitem[{{Yang} {et~al.}(2008){Yang}, {Mo}, \& {van den Bosch}}]{yang08}
{Yang}, X., {Mo}, H.~J., \& {van den Bosch}, F.~C. 2008, \apj, 676, 248

\bibitem[York et al.(2000)]{york00} York, D.~G., et al.\ 2000, 
\aj, 120, 1579 

\bibitem[{{Zheng} {et~al.}(2008){Zheng}, {Zehavi}, {Eisenstein}, {Weinberg}, \&
  {Jing}}]{zheng08}
{Zheng}, Z., {Zehavi}, I., {Eisenstein}, D.~J., {Weinberg}, D.~H., \& {Jing},
  Y. 2008, \apj, submitted (arXiv:0809.1868)

\end{thebibliography}

\end{document}